# Indistinguishable near infra-red single photons from an individual organic molecule

Jean-Baptiste Trebbia, Philippe Tamarat, and Brahim Lounis*

*Centre de Physique Moléculaire Optique et Hertzienne, Université de Bordeaux and CNRS,*

*351 cours de la Libération, Talence, F-33405 (France)*

By using the zero-phonon line emission of an individual organic molecule, we realized a source of indistinguishable single photons in the near infrared. A Hong-Ou-Mandel interference experiment is performed and a two-photon coalescence probability of higher than 50% at 2 K is obtained. The contribution of the temperature-dependent dephasing processes to the two-photon interference contrast is studied. We show that the molecule delivers nearly ideal indistinguishable single photons at the lowest temperatures when the dephasing is nearly lifetime limited. This source is used to generate post-selected polarization-entangled photon pairs, as a test-bench for applications in quantum information.

* blounis@u-bordeaux1.fr

## I. INTRODUCTION

Over the past decade, the quest for photon-based quantum information processing schemes has fueled interest in multiphoton interference phenomena. In particular, linear optics quantum computation and quantum teleportation [1, 2] require the effect of two-photon quantum interference between single photons [3, 4] on a beam-splitter, also called the photon coalescence, whereby indistinguishable photons that arrive simultaneously by two different input ports on a beamsplitter leave both by the same output port [5]. Ideally, a light source that delivers on demand single photons with similar wave packets is desired for these applications. The indistinguishability of the photons generated by single quantum emitters was tested with two-photon quantum interference, for different test-bench systems such as trapped atoms or ions in the gas phase [6-8]. Yet, these systems suffer from weak duty cycles or low light collection efficiency, as well as challenging methods for scaling to large numbers of emitters. Solid-state emitters such as semiconductor quantum dots were used to demonstrate the photon coalescence [9-12] and to produce post-selected entangled states [13], but they usually display residual dephasing and spectral diffusion processes which spoil the coherence properties of the emitted photons. Individual fluorescent molecules in solids are another promising alternative for single photon generation [14, 15]. In comparison to self-assembled quantum dots, the optical coherence lifetime of organic molecules is longer by one order of magnitude, and multi-photon emission is forbidden [16, 17]. At liquid-helium temperatures and for well-chosen fluorophore-matrix systems, dephasing of the transition dipole due to phonons is drastically reduced [18]. Such molecules behave like two-level systems [19] with a fluorescence quantum yield close to unity, thus offering optical properties similar to those of trapped single atoms. Moreover, the photostability of organic molecules trapped in crystalline hosts is excellent at cryogenic temperatures, and allows continuous optical measurements over days [20].

It has been shown that the zero-phonon line (ZPL) emission of a cw-excited single organic molecule produces indistinguishable single photons in the visible domain [14, 15]. In order to test the suitability of a single organic molecule as an essential element for a photon-based quantum logic toolbox, it is crucial to quantify the contribution of dephasing processes to the degradation of the coalescence contrast. In this work, we study the temperature dependence of the two-photon interference contrast, by using the strong near infra-red ZPL emission of single dibenzoterrylene (DBT) molecules embedded in an anthracene (Ac) crystal. This system [21-23] is particularly attractive since the ZPL matches the maximal sensitivity of silicon avalanche photodiodes (APDs) and is suited for propagation in telecommunication fibers. A two-photon interference contrast of 50% is achieved at 2 K and the deviation from the ideal case is attributed to imperfections in the overlap of photon spatial modes, as well as residual coincidences involving background photons. Our molecular source of single indistinguishable photons is also used for the generation of post-selected polarization-entangled photons. Quantum state tomography on photon pairs allows the reconstruction of the polarization density matrix [13, 24].

## II. EXPERIMENTAL SETUP

Single molecules are imaged with a homebuilt scanning confocal optical microscope, based on a solid immersion lens (SIL) used in the Weierstrass configuration and an aspherical lens (see Fig. 1(a) and further details in Ref. [23]). This microscope provides a high collection efficiency of the fluorescence photons (numerical aperture 1.8), as well as a high spatial resolution microscopy of the single molecules (~0.4 $\lambda$). DBT-dopped Ac flakes are mounted on a piezo-scanner, and inserted together with the SIL and the aspherical lens in a helium cryostat. A cw monomode Ti:sapphire laser (767 nm, spectral resolution 1 MHz ) is used to pump individual DBT molecules into a vibronic state (mode at 300 cm$^{-1}$) [see Fig. 1(c)]. After fast (in the picosecond time scale) non-radiative relaxation, fluorescence is emitted from the purely electronic excited state, giving rise to a sharp and intense ZPL (at 785 nm) as well as red-shifted vibrational fluorescence lines, as shown in Fig. 1(d). For DBT molecules in Ac, we found that a relatively large fraction (33%) of the single molecule emission intensity originates from the ZPL at liquid helium temperature [23]. Figure 1(e) shows a zoom of a single DBT molecule emission spectrum on the ZPL and its phonon sideband, recorded with a spectrometer (resolution ~ 90 GHz). Since only the photons arising from the purely electronic ZPL fulfil the indistinguishability requirements for two-photon quantum interference, a severe filtering of the ZPL photons is performed with a band-pass filter (bandwidth 3 nm). Fig. 1 (f) shows the filtered emission spectrum of the same molecule, from which we deduce that ~ 94 % of the filtered photons originate from the ZPL, while other photons are due to residual transmission of the phonon sideband. The fluorescence intensity correlation function measured with ZPL photons displays a strong photon antibunching (zero delay autocorrelation function of ~0.2 in agreement with our signal-to-background ratio of 10, see Fig. 1(g)), which demonstrates single photon emission[23].

The ZPL photons of a single molecule are sent to a Mach-Zender-type setup used to perform two-photon quantum interference, as depicted in Fig. 2. A combination of a quarter-wave plate followed by a half-wave plate and a polarizer is used to set the polarization of the input photons to linear and horizontal. The stream of single photons is then split by a first 50/50 nonpolarizing beamsplitter (NPBS1). Single-mode polarization-maintaining fibers are placed on both arms of the interferometer to facilitate the free-space mode matching of the beams recombined with the second NPBS. The propagation delay between the interferometer arms is set to a value $\Delta t$ ~ 40 ns which is much longer than the maximal coherence time $T_2$ ~ 10 ns of the emitted photons (twice the emitting state lifetime $T_1$ ~ 4.7 ns), to ensure that two photons simultaneously impinging on NPBS2 from different input ports are independent. Two APDs are used in a start-stop configuration to record coincidence counts between the two output ports of NPBS2. A half-wave plate is inserted in one input port of NPBS2 and allows the comparison of coincidence histograms in orthogonal and parallel polarization configurations.

## III. TWO-PHOTON QUANTUM INTERFERENCE

Figures 3 (a) and 3 (b) display the coincidence histograms between output ports (denoted 3 and 4, see Fig. 2), for orthogonal and parallel polarizations of NPBS2 input ports, respectively. These normalized histograms, which give the second order correlation function $g_{34}^{(2)}(\tau)$, are recorded at 2 K for a single DBT molecule continuously excited at an intensity of ~1 MW/cm$^2$ close to optical saturation. The lateral dips at $\Delta t = \pm 40$ $ns$ present in both histograms arise from photon antibunching and reflect the fixed path delay between the interferometer arms. We now compare the relative weights of the central dips of the two histograms, from which we can deduce the coalescence efficiency of the photons. For orthogonal polarizations, the two paths are distinguishable and one expects $g_{34\perp}^{(2)}(0) = 0.5$ due to antibunching for perfectly balanced photon streams on both arms of the interferometer (for unbalanced arms, the central peak becomes deeper at the expense of the lateral ones). In the parallel configuration the two paths are indistinguishable and $g_{34//}^{(2)}(0)$ should ideally reach zero due to antibunching *and* two-photon quantum interference [25, 26]. Indeed, the central dip of $g_{34//}^{(2)}(\tau)$ should display a biexponential rise with one characteristic time related to antibunching and the second to the coherence time of the photon wave-packet. However, because of the limited signal to noise ratio of both coincidence histograms, we use a single exponential rise to fit the central dip of the correlation functions and deduce $g_{34\perp}^{(2)}(0) = 0.59 \pm 0.01$ and $g_{34//}^{(2)}(0) = 0.26 \pm 0.01$. For the lateral dips, we obtain $g_{34\perp}^{(2)}(\pm\Delta t) = 0.76 \pm 0.01$ and $g_{34//}^{(2)}(\pm\Delta t) = 0.76 \pm 0.01$ (close to the ideal value of 0.75) which ensures that the photon streams on the two arms of the interferometer are balanced within 5 %. The value of $g_{34//}^{(2)}(0)$ readily demonstrates two-photon coalescence on NPBS2. The two-photon interference contrast, which can be defined as $C(\tau) = \left[g_{34\perp}^{(2)}(\tau) - g_{34//}^{(2)}(\tau)\right]/g_{34\perp}^{(2)}(\tau)$, is shown in Fig. 3 (c), and takes its maximum value $C(0) = 0.51$ which is close to the typical values obtained with self-assembled quantum dots in microcavity structures [9]. The characteristic decay time of $C(\tau)$, which is half the optical coherence lifetime [26], is in broad agreement with that expected from the measured ZPL width (~ 40 MHz).

The deviation of the two-photon interference contrast $C(0)$ from its ideal unity value has two origins. A main contribution is due to imperfections in the spatial overlap of the output beams on NPBS2. The wavefront mismatch and misalignment is estimated from interference patterns of a laser beam tuned to the same wavelength as the molecular ZPL and sent into the interferometer. Residual phase shifts found in the section of the output beams (misalignment and mismatch of λ/2 over a beam diameter 22 mm) are expected to reduce the two-photon interference contrast by ~30%. Another contribution of residual zero delay coincidences is due to events involving the background. A signal-to-background of ~20 is indeed measured in confocal images of the studied single DBT molecules, and leads to background coincidence contribution of ~ 10%. Finally, the binwidth of coincidence histograms reduces the coalescence dip by about ~ 10%. This analysis suggests that dephasing of the molecular transition dipole due to phonons in the crystalline matrix is negligible at 2 K, enabling a single DBT molecule to deliver indistinguishable single photons.

## IV. TEMPERATURE-DEPENDENCE OF THE COALESCENCE CONTRAST

A quantitative study of the photon coalescence degradation by dephasing processes is required to verify the potentialities of this single photon source in quantum information processing. It was previously shown that artificial spectral broadening of a single molecule ZPL with AC Stark effect leads to a reduced probability of coalescence [15]. Rather, we focused our interest on the contribution of phonons to dephasing processes, since the host matrix is cooled down to a finite temperature. Theoretically, dephasing is expected to reduce the probability of coalescence to $T_2/(2T_1)$ [27]. We investigate the dependence of the coalescence efficiency on the optical coherence time by varying the temperature of the host matrix. The probability of coalescence is deduced from the coincidence histograms as $(A_{//} - A_{\perp})/A_{\perp}$, where $A_{//}$ and $A_{\perp}$ are the areas of the central dip in the parallel and the orthogonal polarization configurations, respectively. As displayed in Fig. 4, this probability drops by a factor of ten upon increasing the temperature from 2 K to 5 K. The dephasing time $T_2$ is derived from the ZPL width measured from the resonant fluorescence excitation spectrum [18]. As shown in the lower inset of Fig. 4, temperature-dependent dephasing processes result in a drastic homogenous broadening of the molecular resonance. In crystalline matrices of small, rigid molecules at low temperatures, dephasing is often induced by an optical phonon mode (due to libration of the guest molecule), and the temperature-dependent linewidth is reproduced with an Arrhenius law [20, 28]. For this molecule, the linewidth is 32 MHz in the low temperature regime and the activation temperature is 16 K (see upper inset of Fig. 4), in agreement with previous measurements on this fluorophore-matrix system [28]. Using the evolution of the temperature-dependent coherence lifetime allows to reproduce the variation of the coalescence probability with temperature [see solid curve of Fig. 4], without adjusting parameter except a global scale factor which takes into account the temperature-independent sources of coalescence degradation described above. For temperatures T ≤ 2 K, dephasing becomes negligible ($T_2/(2T_1)$ approaches unity) and single DBT molecules are nearly ideal sources of single indistinguishable photons.

## V. POST-SELECTED POLARIZATION ENTANGLEMENT

In the view of potential applications in linear optics quantum computation, it is interesting to generate polarization entangled-photons with this test-bench single photon source. For instance, when colliding photons with orthogonal polarizations at the input ports of NPBS2, quantum interference ensures that photons simultaneously detected at different output ports should be entangled in polarization. More precisely, since the coincidence measurements do not detect photon pairs arriving on the same detector, the output polarization state $(|H_3V_3,0_4\rangle + |H_3, V_4\rangle - |V_3,H_4\rangle - |0_3,H_4V_4\rangle)/2$, is reduced to the post-selected Bell state $(|H_3, V_4\rangle - |V_3,H_4\rangle)/\sqrt{2}$ obtained with a probability of 1/2, where H and V specify the polarization (horizontal or vertical) of an output spatial mode. The generation of entangled photons was performed with the same experimental setup, setting H and V polarizations on the paths of the Mach-Zender interferometer. A combination of a quarter-wave plate followed by a half-wave plate and a polarizer was placed before each APD, in order to characterize the generated polarization-entangled state with quantum state tomography [see Fig. 5 (a)]. Sixteen measurements were performed with different analyser settings, including crossed diagonal linear polarizations and opposite helicity circular polarizations which directly test the non-separability of the photon pairs states [29]. The values of $g^{(2)}_{34}(0)$ were measured in all analyser configurations, and corrected from coincidences due to background accidental events which are determined with antibunching measurements. Since background is not polarized, the subtracted background contribution was identical in all correlation histograms. The latter were normalized to their values at large delay times. The values of $g^{(2)}_{34}(0)$ were then used to reconstruct the density matrix of the two-photon polarization state with the maximum likelihood algorithm [24]. We obtain the following density matrix

$$\rho = \begin{bmatrix} 0.02 & 0.02 + 0.08i & -0.07 - 0.07i & -0.01 + 0.01i \\ 0.02 - 0.08i & 0.43 & -0.23 + 0.1i & 0.08 + 0.08i \\ -0.07 + 0.07i & -0.23 - 0.1i & 0.50 & -0.02 + 0.02i \\ -0.01 - 0.01i & 0.08 - 0.08i & -0.02 - 0.02i & 0.05 \end{bmatrix}$$

in the $(H/V) \otimes (H/V)$ basis, where the ideal matrix is $\rho_{Ideal} = \frac{1}{2}\begin{bmatrix} 0 & 0 & 0 & 0 \\ 0 & 1 & -1 & 0 \\ 0 & -1 & 1 & 0 \\ 0 & 0 & 0 & 0 \end{bmatrix}$. The real and imaginary components of $\rho$ are represented in Fig. 5 (b) in absolute values for clarity, and compared to the ideal case displayed in Fig. 5 (c). The diagonal elements $\rho_{HV,HV} = 0.43$ and $\rho_{HV,HV} = 0.50$ are close to the ideal values 0.5. The real part of the off-diagonal elements $\rho^{real}_{VH,HV} = -0.23$ and $\rho^{real}_{HV,VH} = -0.23$ is close to that expected when one takes into account the reduction of the ideal value by ~50% due to the reduced contrast of two-photon interference (i.e. mismatch in the recombined photons spatial modes) [13]. Nevertheless, the quantum inseparability which shows up in the density matrix can be quantified with the Peres criterion [30], based on the sign of the eigenvalues of a matrix obtained by partial transposition of the density matrix. Here the Peres criterion takes the value -0.25, a value of -1 meaning maximum entanglement.

## VI. CONCLUSION

In summary, we showed that the zero-phonon line emission of a single DBT molecule is well suited for high visibilty two-photon quantum interference. The generation of polarization-entangled photon pairs from a molecular single photon source is also demonstrated. The study of the coalescence efficiency degradation by temperature-dependent dephasing processes indicates that a single DBT molecule trapped in an anthracene crystal is a source of nearly ideal single indistinguishable photons in the near infra-red. This system can therefore be of direct interest for implementation in quantum information processing schemes. Further developments will aim at increasing the emission rate of indistinguishable photons and improving the photons spatial mode overlap, e.g. by coupling the molecule to a resonant microcavity and using photonic waveguide circuits.

## ACKNOWLEDGMENTS

This work was funded by the Agence Nationale de la Recherche (Program PNANO), Région Aquitaine, and the European Research Council. We thank H. Ruf for his help at the early stages of these experiments.

**Figures**

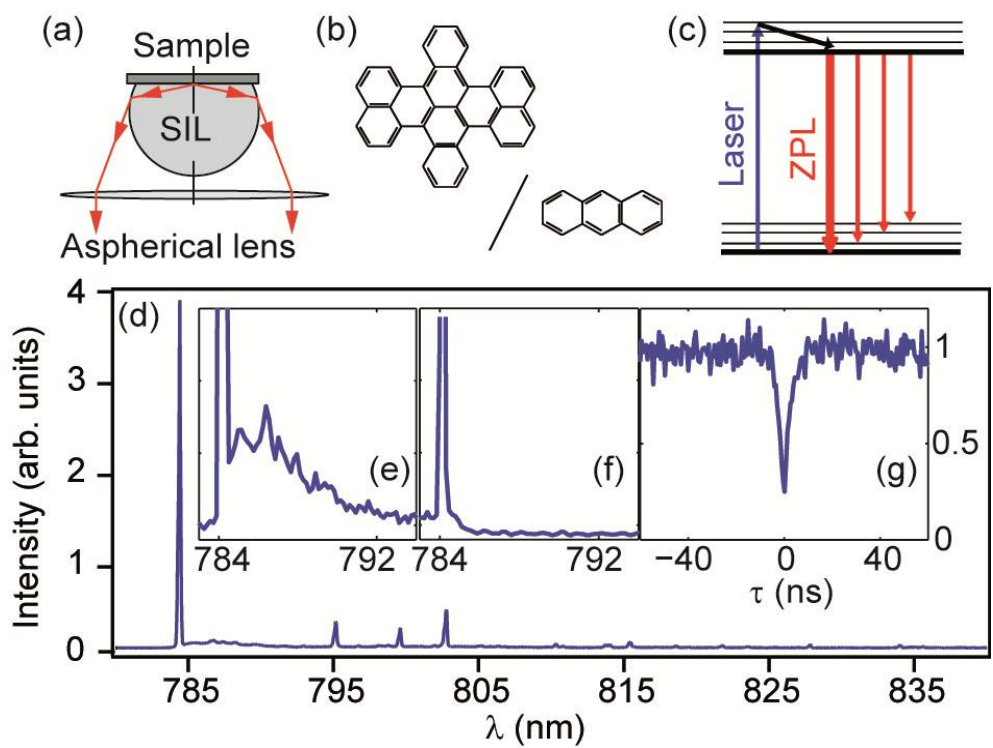

Figure 1: (Color online) (a) Scheme of our home-built scanning confocal microscope which combines a SIL, an aspherical lens an a piezo-scanner. (b) Chemical structure of the DBT/Ac system. (c) Energy level diagram of a DBT molecule, pumped into a vibronic level. (d) Fluorescence photons are emitted on the purely electronic transition (ZPL), as well as red-shifted vibrational transitions. A total fluorescence saturated count rate of ~0.6 million counts per second is achieved with this system. (e) Emission spectrum of a single DBT molecule at 2 K, displaying the sharp and intense ZPL and its red-shifted phonon sideband. (f) Emission spectrum of the same molecule after ZPL filtering through a 3 nm band-pass filter. 94% of the transmitted photons stem from the ZPL which has linewidth of ~ 37 MHz at 2 K. (g) Fluorescence intensity autocorrelation function measured with ZPL photons emitted by a single DBT molecule, using a Hanbury-Brown and Twiss setup.

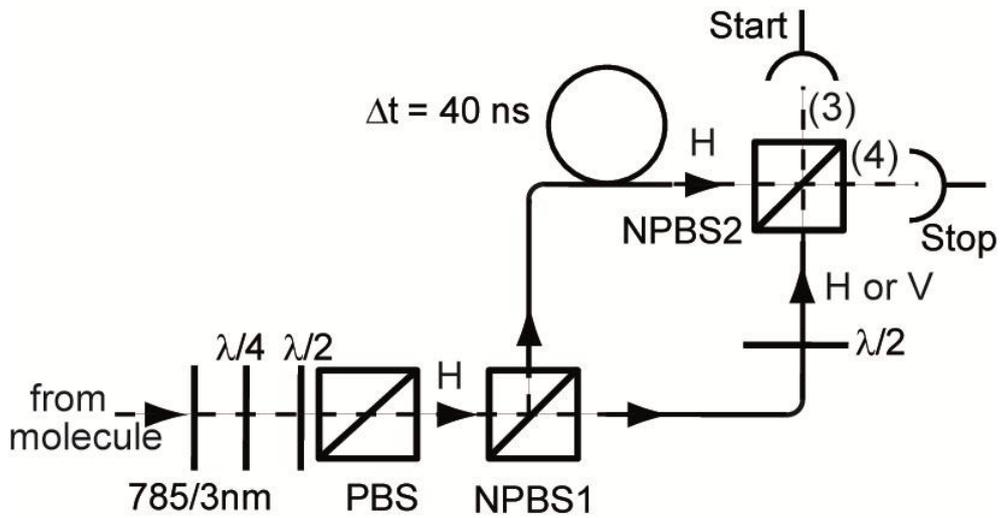

Figure 2: Scheme of the experimental setup. The ZPL emission from a single molecule is filtered, set linearly and horizontally (H) polarized, and sent to a Mach-Zender-type interferometer composed of two 50/50 non polarizing beamsplitters (NPBS). Single mode polarizing-maintaining fibers introduce a 40 ns delay between both arms and ensure a satisfying spatial mode overlap of the output photon streams. The histogram of time delays between start and stop clicks of the APDs is recorded for parallel (H,H) or orthogonal (H,V) polarizations of the photons colliding on NPBS2.

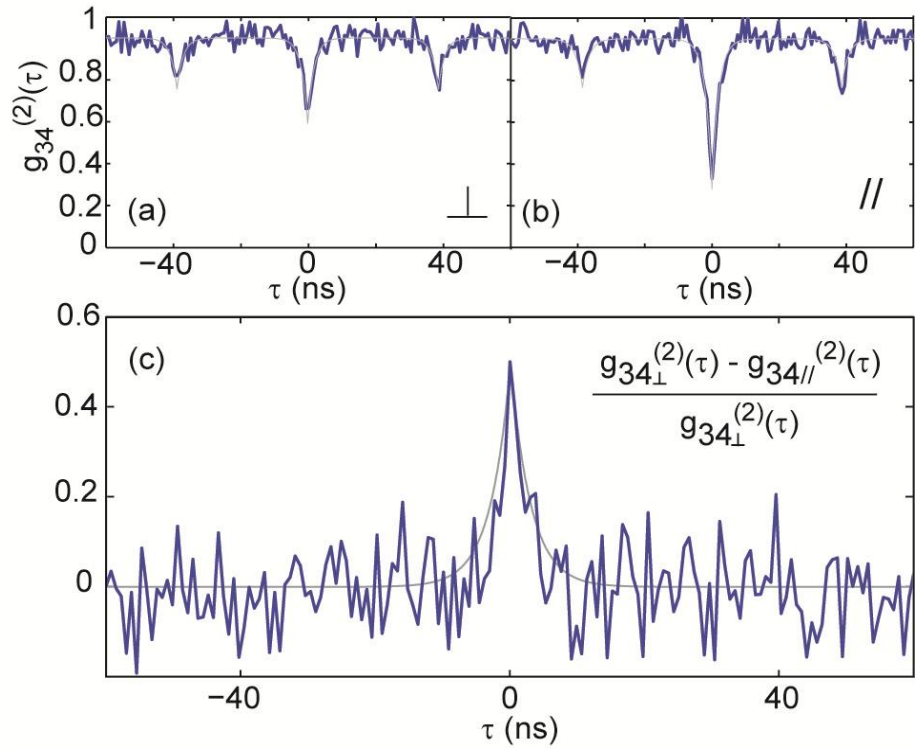

Figure 3 : (Color online) Coincidence histograms recorded for a single DBT molecule at 2 K, for orthogonal (a) and parallel (b) polarizations of the interferometer arms. Solid curves are fits of the temporal coincidence profiles with a threefold exponential rise function. (c) Normalized difference between the coincidence histograms in parallel and orthogonal polarizations. The integration time is ~2 hours for each histogram, the count rate is ~20 kHz per channel, and the binwidth is 760 ps.

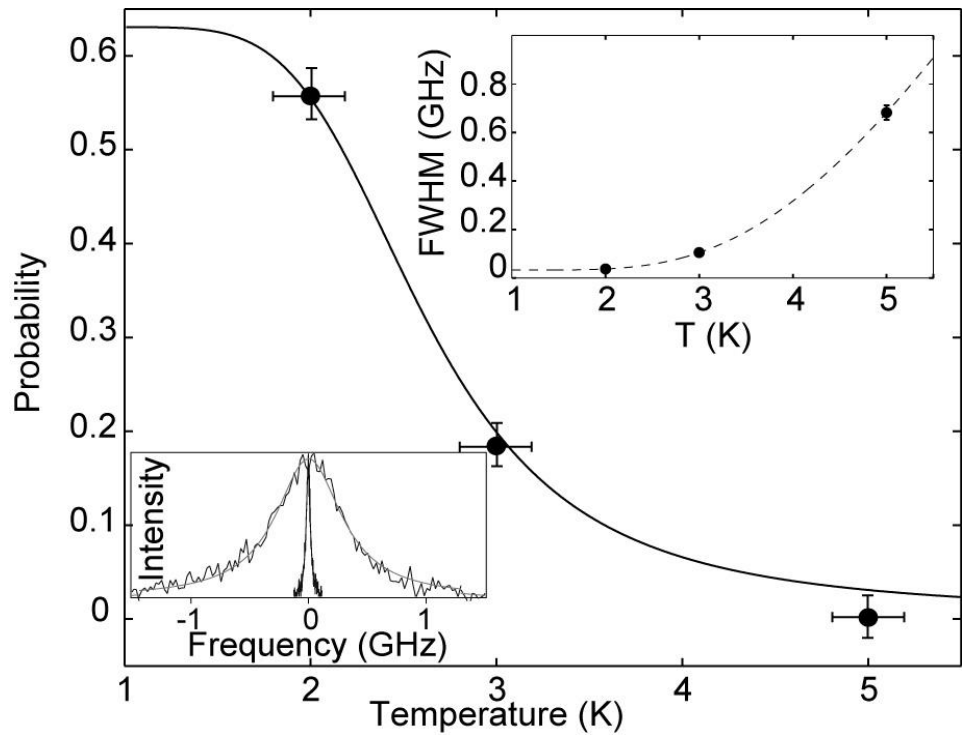

Figure 4: Temperature dependence of the coalescence probability for a single DBT molecule. The solid curve gives the evolution of the theoretical probability $T_2/(2T_1)$, corrected with a global scaling factor due to imperfections of the spatial modes overlap. $T_2$ is derived from the ZPL linewidth. Lower inset: Excitation spectra of the molecule at 2 K and 5 K. Upper inset: The dotted line is an Arrhenius law guiding the temperature dependence of the ZPL width (plain circles).

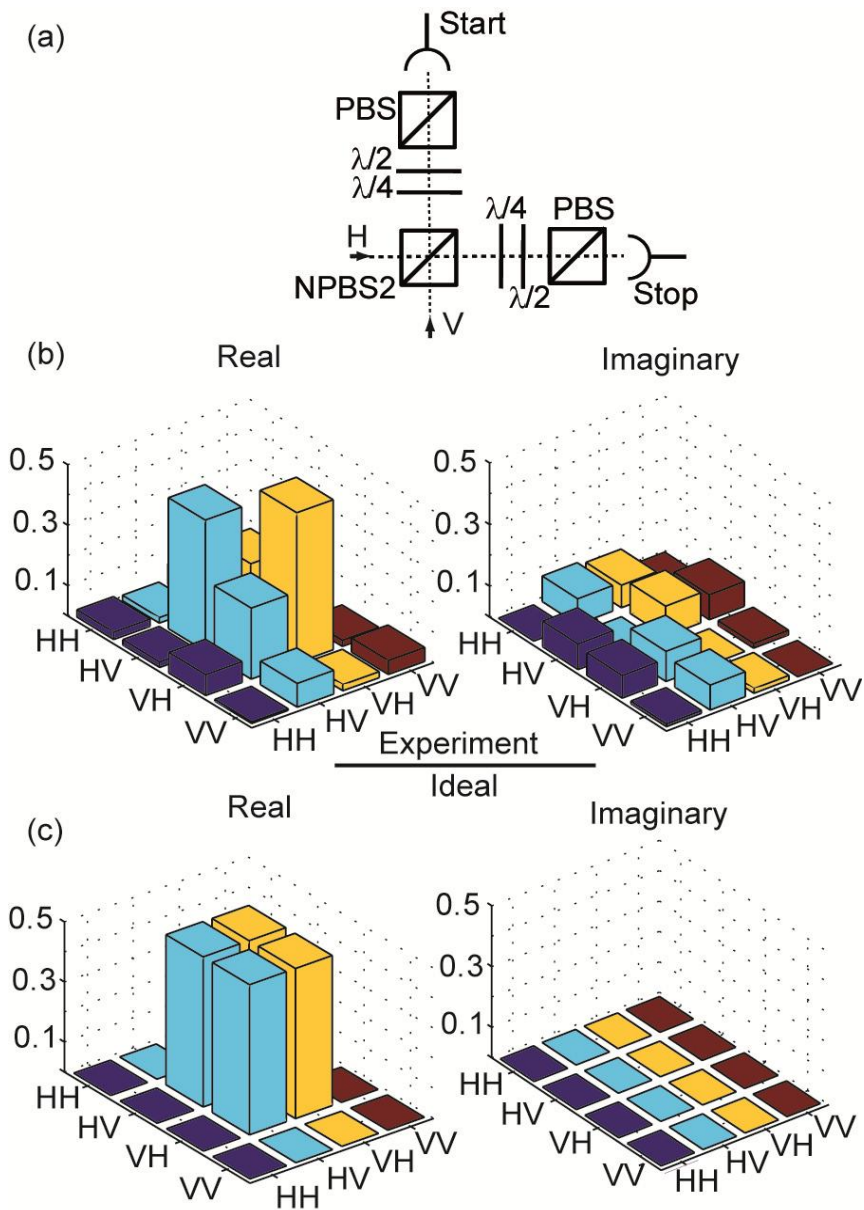

Figure 5 : (Color online) (a) Setup of polarization analysis added to the experimental setup shown in Fig. 2. (b) Reconstructed density matrix of the two-photon polarization state. Real and imaginary components are represented in absolute values, and compared to the ideal density matrix (c).